\documentclass[twocolumn]{jpsj2} 
\title{Magnetic phase diagram of three-dimensional diluted Ising
antiferromagnet Ni$_{0.8}$Mg$_{0.2}$(OH)$_{2}$}

\author{Masatsugu Suzuki$^{1}$\thanks{E-mail address: suzuki@binghamton.edu},
Itsuko S. Suzuki$^{1}$\thanks{E-mail address: itsuko@binghamton.edu},
Tedmann M. Onyango$^{1}$ and
Toshiaki Enoki$^{2}$}

\inst{$^{1}$Department of Physics, State University of New York at Binghamton,
Binghamton, New York 13902-6016, USA \\
$^{2}$Department of Chemistry, Tokyo Institute of Technology,
Meguro-ku, Tokyo 152-8551}

\abst{$H$-$T$ diagram of 3D diluted Ising antiferromagnet
Ni$_{c}$Mg$_{1-c}$(OH)$_{2}$ with $c$ = 0.8 has been determined from
measurements of SQUID DC magnetization and AC magnetic susceptibility. At
$H$ = 0, this compound undergoes two magnetic phase transitions: an
antiferromagnetic (AF) transition at the N\'{e}el temperature $T_{N}$ (=
20.7 K) and a reentrant spin glass (RSG) transition at $T_{RSG}$
($\approx$ 6 K). The $H$-$T$ diagram consists of the RSG, spin glass (SG),
and AF phases. These phases meet a multicritical point $P_{m}$ ($H_{m}$ = 42 kOe,
$T_{m}$ = 5.6 K). The irreversibility of susceptibility
defined by $\delta$ (= $\chi_{FC} - \chi_{ZFC}$) shows a negative local
minimum for 10 $\leq H \leq$ 35 kOe, suggesting the existence of possible
glassy phase in the AF phase. A broad peak in $\delta$ and $\chi^{\prime
\prime}$ at $H \geq$ 20 kOe for $T_{N}(c=0.8,H) \leq T \leq T_{N}(c=1,H=0)$
(= 26.4 K) suggests the existence of the Griffiths phase.}

\kword{spin glass, Griffiths phase, random magnet}

\begin{document}
\maketitle

\section{\label{intro}Introduction}
The static and dynamics of Ising random spin systems present some of the
most intriguing problems in magnetism \cite{Young1998}. There are two types
of spin ordering in Ising random spin systems that have attracted
continuing interest for two decades: Ising spin glass (SG) and random-field
Ising model (RFIM). The ordered phase of both cases is characterized by
the irreversibility and the metastability due to multiple minima of the
free-energy surface. The spin glass behavior is realized in random Ising
spin systems with competing exchange interaction. The RFIM problems have
been extensively discussed from a theoretical view
point \cite{Young1998,Imry1975,Fishman1979}. Experimentally it is impossible
to create random fields on a microscopic scale. However, the RFIM can be
realized in an diluted Ising antiferromagnet in an uniform external
magnetic field ($H$) \cite{Fishman1979}. The random-field strength is
predicted to be proportional to $H$.

In this paper we study the nature of glassy phases in a three-dimensional
(3D) diluted Ising antiferromagnet Ni$_{c}$Mg$_{1-c}$(OH)$_{2}$ with $c$ =
0.8 in the presence of $H$. This compound shows two types of transition
at $H$ = 0:
(i) the transition between the antiferromagnetic (AF) and paramagnetic (PM)
phases at a N\'{e}el temperature $T_{N}$ (= 20.7 K) and (ii) the transition
between the AF and reentrant spin glass (RSG) phases at a transition
temperature $T_{RSG}$ ($\approx$ 6 K) \cite{Suzuki2000}.
The present work has been
motivated by the following three theoretical predictions on various types
of glassy phases. The first is the prediction for the spin glass phase
related to the AF system with RSG phase in an uniform $H$.
The magnetic phase ($H$-$T$) diagram of an AF-based RSG (AF-RSG) system has
been theoretically studied by Takayama \cite{Takayama1988} and Fyodorov et
al. \cite{Fyodorov1990} based on a molecular field theory (see \S
\ref{back}). These theories predict that the $H$-$T$ diagram consists of
four phases, an AF phase, a RSG phase, a pure SG phase, and a PM phase.
These four phases meet at a new multicritical point. An AF-RSG transition
temperature $T_{RSG}(H)$ tends to increase with increasing $H$. The
extrapolation of the phase boundary between the AF phase and PM phase to
the low-$T$ side separates the RSG phase and SG phase, where the SG phase
is in the higher-$H$ side. The second is the prediction for the glassy
phase in a 3D diluted Ising antiferromagnet under an uniform $H$ as a
realization of the RFIM. It has been theoretically predicted that a glassy
phase exists between the PM phase and AF phase in the $H$-$T$ phase diagram
of the 3D diluted Ising antiferromagnet in an uniform field
$H$ \cite{Soukoulis1983,Ro1985,Soukoulis1985,Grest1986,Almeida1987}.
There are two-types of states at the same $H$ and $T$: zero-field (ZFC) and
field-cooled (FC) states. The FC state is obtained upon cooling of the
system at constant $H$ and the ZFC state is obtained by first cooling at
$H$ = 0 and then applying $H$ and reheating. The ZFC state (which has long
range order) has a lower free energy in the AF phase, while the FC state
(the multi-domain state) has a lower free energy in the glassy
phase \cite{Yoshizawa1984}. However, it is not clear whether the glassy
phase is regarded as a truly separate phase or simply as having long
relaxation times due to domain-wall pinning. Experimentally it is expected
that the ZFC magnetization $M_{ZFC}$ is larger than the FC magnetization
$M_{FC}$ in the glass phase ($H > H_{crit}$) and that the opposite
inequality holds for $H < H_{crit}$ \cite{Ro1985}, where the crossover line
between the glassy phase and the AF phase is denoted by $H$ =
$H_{crit}(T)$. This may be an useful experimental signature of the onset
of the glassy phase. The third is a so-called Griffiths phase
conjecture \cite{Griffiths1969} which is based on the idea of local phase
transitions in a diluted system due to the finite probability of
arbitrarily large pure and differently diluted clusters. The Griffiths
phase is the phase of the slowly fluctuating spins that exists in the
paramagnetic region $T_{N}(c,H) \leq T \leq T_{N}(c=1,H)$, where
$T_{N}(c,H)$ is the N\'{e}el temperature of the system with the
concentration $c$.

As far as we know, there has been at least one report in a 3D diluted
Ising antiferromagnet (Fe$_{0.552}$Mg$_{0.448}$Cl$_{2}$) \cite{Gelard1983},
which supports the first prediction. At $H$ = 0
this compound undergoes a PM-AF transition at $T_{N}$ (= 7.5 K) and AF-RSG
transition at $T_{RSG}$ ($\approx$ 3.0 K). The RSG phase is a mixed phase
where SG behavior and AF long range-order coexist \cite{Wong1985}. An
uniform magnetic field $H$ along the $c$ axis enlarges the domain of
existence of the RSG phase in the ($T$, $H$) plane for $H <$ 5 kOe:
$T_{RSG}(H)$ increases with increasing $H$ and meets with $T_{N}(H)$ at a
multicritical point. Such an increase of $T_{RSG}$ may give an
experimental evidence for the random field (RF) effect \cite{Gelard1983}.
The existence of the Griffiths phase has been reported in pure
FeCl$_{2}$ \cite{Binek1994} and Fe$_{0.47}$Zn$_{0.53}$F$_{2}$
\cite{Binek1995}. Binek et al. \cite{Binek1994,Binek1995} have shown
that there exist domain-like antiferromagnetic correlations in the
temperature range $T_{N}(c,H) \leq T \leq T_{N}(c=1,H=0)$, when $H$ is
applied along the $c$ axis. As far as we know, there has been no report on
the Griffiths phase in Fe$_{c}$Mg$_{1-c}$Cl$_{2}$ with 0.5 $\leq c \leq$
0.61. We note that the existence of the cluster-ordered Griffiths phase in
Ni$_{c}$Mg$_{1-c}$(OH)$_{2}$ has been claimed by Deguchi et
al. \cite{Deguchi1998} from their SQUID DC magnetization measurement and by
Zenmyo et al. \cite{Zenmyo2000} from proton nuclear magnetic resonance.

In this paper we have determined the $H$-$T$ phase diagram of
Ni$_{c}$Mg$_{1-c}$(OH)$_{2}$ with $c$ = 0.8
from the measurements of DC magnetic susceptibility and AC magnetic
susceptibility (1.9 $\leq T \leq$ 30 K, 0 $\leq H \leq$ 48 kOe, 0.05 $\leq
f \leq$ 1000 Hz). We show that for low $H$ (0 $\leq H \leq$ 3 kOe) the
difference $\delta$ (= $\chi_{FC}$ - $\chi_{ZFC}$) is positive for any $T$
and exhibits two step-like increases at different temperatures with
decreasing $T$: a weak irreversibility at $T$ just below $T_{N}(H)$ due to
a possible RF effect and a strong irreversibility at $T_{RSG}(H)$ due to
the RSG phase. In contrast, $\delta$ for 10 $\leq H \leq$ 48 kOe is
negative below $T_{N}(H=0)$, suggesting the possible glassy phase between
the PM and AF phases as a result of RFIM. We find the strong $H$ dependence
of the absorption $\chi^{\prime \prime}$ at low $T$, showing a peak at $H
\approx$ 42 kOe, suggesting the separation between the RSG and SG phase at
low $T$. We also find the Griffiths phase in Ni$_{0.8}$Mg$_{0.2}$(OH)$_{2}$.
The absorption $\chi^{\prime \prime}$ shows a very broad
peak at $T$ between $T_{N}(c=0.8,H)$ and $T_{N}(c=1,H=0)$ (= 26.4 K). The frequency
dependence of $\chi^{\prime \prime}$ vs $T$ is also examined at $H$ = 30
and 40 kOe to determine the average relaxation time $\tau_{G}$.

\section{\label{back}Background: $H$-$T$ phase diagram for the AF based RSG systems}

\begin{figure}
\begin{center}
\includegraphics[width=8.0cm]{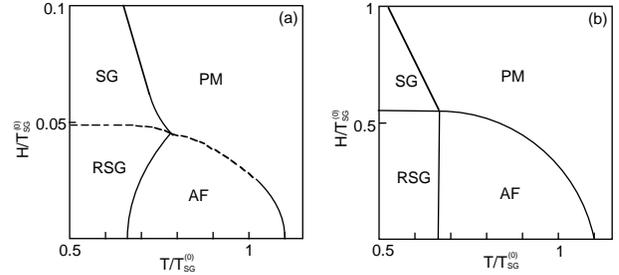}%
\end{center}
\caption{The $H$-$T$ diagram predicted from the mean-field
theory \cite{Takayama1988} for (a) an enhanced AF-RSG system with $r$ =
$r_{F}$ = 5 and $r_{N}$ = 1.1, and (b) a simple AF-RSG system with $r$ =
$r_{F}$ = 0 and $r_{N}$ = 1.1. The dashed line in (a) is of the first
order. The definition of $r$, $r_{F}$, $r_{N}$, and $T_{SG}^{(0)}$ is
given in the text.}
\label{fig:one}
\end{figure}

Here we present a simple review on the theory of the $H$-$T$ diagram for an
AF based reentrant spin glass (AFRSG). According
to Takayama \cite{Takayama1988}, the system consists of two Ising spin glass
sublattices \{$S_{1i}$\} and \{$S_{i2}$\}, which are coupled by the
intra-sublattice exchange interaction $I_{ij}$ and are coupled
antiferromagnetically on average through the inter-sublattice exchange
interactions $J_{ij}$. These interactions are assumed to be independent
Gaussian random variables with the distribution of the mean value $I_{0}/N$
and $-J_{0}/N$ and variances $I^{2}/N$ and $J^{2}/N$, where $I_{0}>$ 0 and
$J_{0}>$ 0, $I_{0}=r_{F}J_{0}$, $I=rJ$, and $N$ is the number of spins in
each sublattice. Figures \ref{fig:one}(a) and (b) show the $H$-$T$ diagram
of the AFRSG system with (a) $r=r_{F}$ = 5 and $r_{N}$ = 1.1 and (b)
$r=r_{F}$ = 0 and $r_{N}$ = 1.1, respectively, where
$T_{N}^{(0)}=(1+r_{F})J_{0}$ and $T_{SG}^{(0)}=(1+r^{2})^{1/2}J$, and
$r_{N}=T_{N}^{(0)}/T_{SG}^{(0)}$. There are four phases : PM, AF, RSG, and
pure SG phases. In the case of (a), there is a crossover from the second
order transition to the first order transition on the boundary between the
PM and AF phases. The boundary between the SG and RSG phases is also of the
first order. The AF-RSG transition temperature $T_{RSG}$ increases with
increasing $H$. The case (b) ($r=r_{F}$ = 0) corresponds to the model
discussed by Fyodorov et al. \cite{Fyodorov1990}.

\section{\label{exp}Experimental procedure}
Powdered sample of Ni$_{0.8}$Mg$_{0.2}$(OH)$_{2}$
used in the present work
was the same as those used in the previous work \cite{Suzuki2000}. The
detail of sample preparation was reported before \cite{Enoki1975a}. The
DC magnetization and AC magnetic susceptibility were measured using a
SQUID (superconducting quantum interference device) magnetometer (Quantum
Design, MPMS XL-5).

(i) \textit{SQUID DC magnetization}. Before setting up a sample at 298 K,
a remanent magnetic field in a superconducting magnet was reduced to one
less than 3 mOe using an ultra low field capability option of the SQUID
magnetometer. For convenience, hereafter this remanent field is denoted as
the state of $H$ = 0. The sample was cooled from 298 to 1.9 K at $H$ = 0.
Then an external magnetic field $H$ (1 Oe $\leq H \leq$ 48 kOe) was applied
at 1.9 K. A zero-field cooled magnetization ($M_{ZFC}$) was measured with
increasing $T$ from 1.9 to 30 K. After the ZFC measurement, the sample was
heated and kept at 100 K for 20 minutes. It was again cooled to 30 K in
the presence of the same $H$. A field-cooled magnetization ($M_{FC}$) was
measured with decreasing $T$ from 30 to 1.9 K.

(ii) \textit{SQUID AC magnetic susceptibility}. The frequency ($f$)
dependence of $\chi^{\prime}$ and $\chi^{\prime \prime}$ was measured at
various $T$. The sample was cooled from 298 to 1.9 K at $H$ = 0. An
external magnetic field ($H$ = 30 and 40 kOe) was applied at 1.9 K. Then
both $\chi^{\prime}$ and $\chi^{\prime \prime}$ were simultaneously
measured as a function of frequency (0.1 Hz $\leq f \leq$ 100 Hz) at fixed
$T$. $T$ is increased by $\Delta T$ = 0.5 K after each frequency scan.
The amplitude of AC magnetic field $h$ was 3 Oe. The $T$ dependence of
$\chi^{\prime}$ and $\chi^{\prime \prime}$ was also measured in the
presence of various $H$, where $f$ = 1 Hz and $h$ = 3 Oe. The sample was
cooled from 298 to 1.9 K in $H$ = 0. Then $\chi^{\prime}$ and
$\chi^{\prime \prime}$ were simultaneously measured with increasing $T$
from 1.9 to 30 K in the presence of fixed $H$. After each $T$ scan, $H$
was changed at 30 K and $T$ was decreased from 30 to 1.9 K. Then the
measurement was repeated at the same $H$.

\section{\label{result}Result}
\subsection{\label{resultA}$\chi_{ZFC}$ and $\chi_{FC}$}

\begin{figure}
\begin{center}
\includegraphics[width=7.5cm]{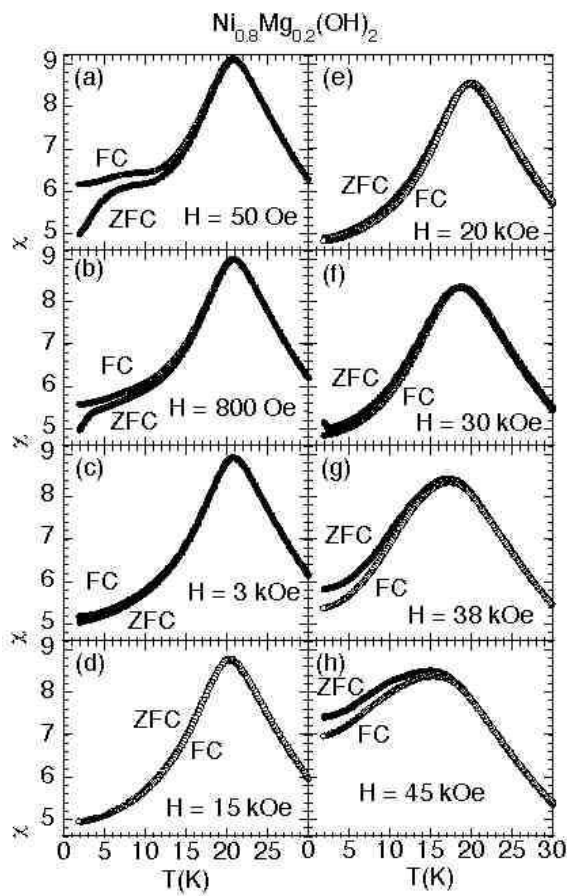}%
\end{center}
\caption{(a) - (h) $T$ dependence of $\chi_{ZFC}$ (=
$M_{ZFC}/H$) ({\Large $\bullet$}) and $\chi_{FC}$ (= $M_{FC}/H$)
({\Large $\circ$})
for Ni$_{0.8}$Mg$_{0.2}$(OH)$_{2}$ at various $H$ (50 Oe $\leq
H \leq$ 45 kOe).}
\label{fig:two}
\end{figure}

\begin{figure}
\begin{center}
\includegraphics[width=7.5cm]{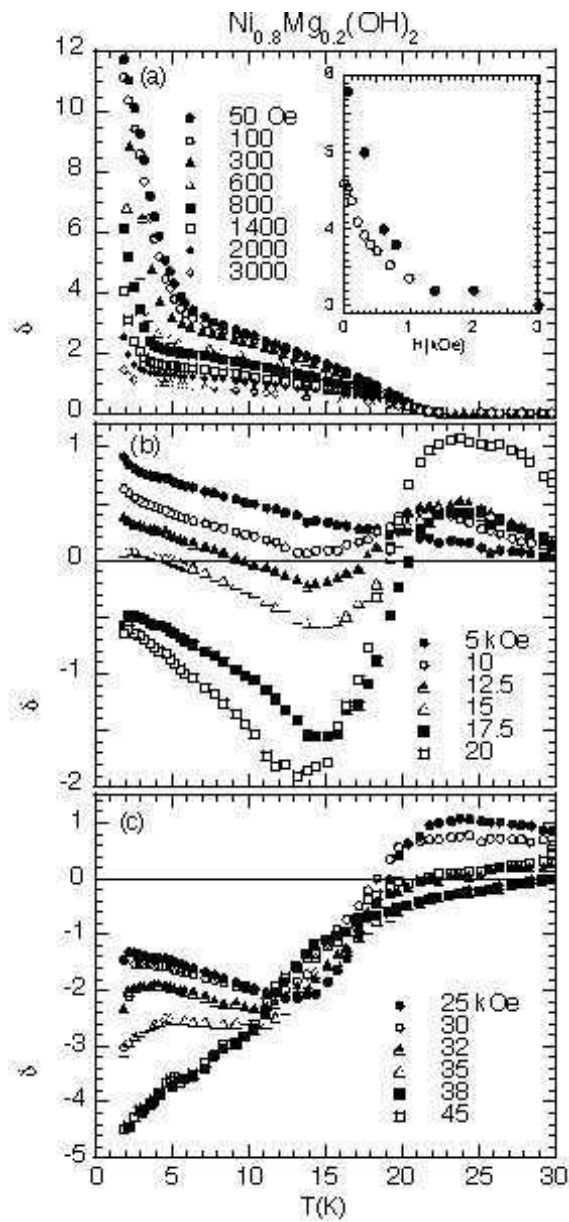}%
\end{center}
\caption{(a)-(c) $T$ dependence of $\delta$ (=
$\chi_{FC}-\chi_{ZFC}$) at various $H$ (50 Oe $\leq H \leq$ 45
kOe). The inset of (a) shows the $H$ dependence of $T_{RSG}$.
$T_{RSG}$ is defined as a temperature at which $\delta$
shows a strong irreversibility ({\Large $\bullet$}) or $\chi^{\prime \prime}$ has
a peak ({\Large $\circ$}).}
\label{fig:three}
\end{figure}

Figure \ref{fig:two} shows the $T$ dependence of $\chi_{ZFC}$ (=
$M_{ZFC}/H$) and $\chi_{FC}$ (= $M_{FC}/H$) for Ni$_{0.8}$Mg$_{0.2}$(OH)$_{2}$
at various $H$ (50
Oe $\leq H \leq$ 45 kOe). Figure \ref{fig:three} shows the $T$ dependence
of the difference $\delta$ ( = $\chi_{FC}-\chi_{ZFC}$) at various $H$. At
$H$ = 50 Oe both $\chi_{ZFC}$ and $\chi_{FC}$ exhibit a peak at 20.8 K. The
deviation of $\chi_{ZFC}$ from $\chi_{FC}$ starts to occur below 22 K,
indicating the positive sign of $\delta$. Note that $\delta$ increases in
two steps with decreasing $T$: a weak change near 22 K and a strong change
below 5.8 K. As will be discussed in \S \ref{disD}, the strong
irreversibility is related to the RSG ordering \cite{Suzuki2000}, while the
weak irreversibility is due to the RF effect. For simplicity $T_{RSG}$ and
$T_{RF}$ are defined as temperatures at which $\delta$ starts to show
strong and weak irreversibility, respectively. For 100 Oe $\leq H <$ 3
kOe, similar step-like changes in $\delta$ are also observed at $T_{RSG}$
and $T_{RF}$. Here $T_{RSG}$ decreases with increasing $H$ ($T_{RSG}$ =
5.8 K at 50 Oe and 3.0 K at 3 kOe), while $T_{RF}$ seems to be independent
of $H$ ($T_{RF}$ = 22.0 - 22.4 K) and is a little higher than the N\'{e}el
temperature $T_{N}$ ($c$ = 0.8, $H$ = 0) (= 20.7 K). The $H$ dependence of
$T_{RSG}$ thus obtained is plotted in the inset of Fig.
\ref{fig:three}(a).

\begin{figure}
\begin{center}
\includegraphics[width=7.5cm]{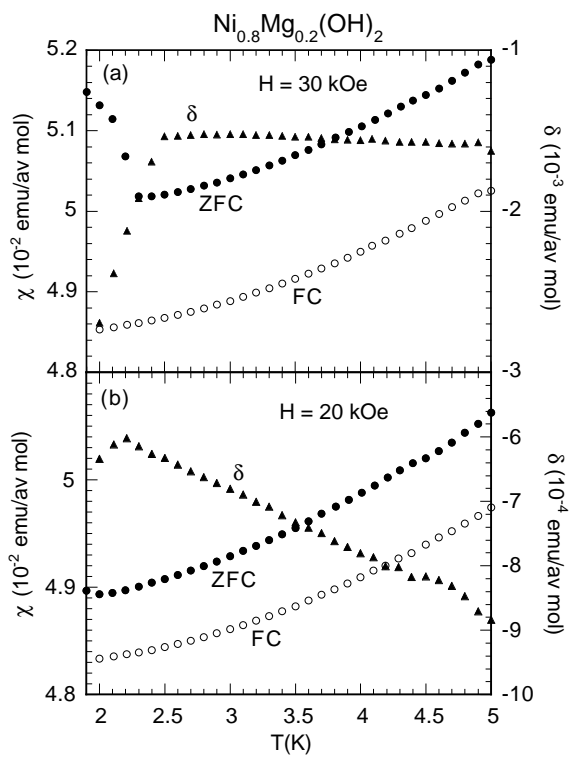}%
\end{center}
\caption{(a) and (b) $T$ dependence of $\chi_{ZFC}$, $\chi_{FC}$,
and $\delta$ at $H = 30$ and 20 kOe around $T_{\delta}$.}
\label{fig:fournew}
\end{figure}

The magnetic behavior for $H \geq$ 10 kOe is rather different from
that for $H \leq$ 3 kOe. At $H$ = 20 kOe, $\chi_{ZFC}$ and
$\chi_{FC}$ exhibit a peak at 19.8 and 20.0 K, respectively. The
sign of $\delta$ changes from positive to negative around 18.9 K
as $T$ decreases, implying that the value of $\chi_{ZFC}$ is
larger than that of $\chi_{FC}$. The difference $\delta$ shows a
negative local minimum at a characteristic temperature $T_{min}$
(= 13.1 K). At $H$ = 30 kOe, $\chi_{ZFC}$ and $\chi_{FC}$ exhibit
a peak at 18.5 and 18.8 K, respectively. The sign of $\delta$
changes from positive to negative at 18.3 K as $T$ decreases. The
difference $\delta$ shows a negative local minimum at 11.7 K. A
drastic decrease in $\delta$ with decreasing $T$ is seen below a
characteristic temperature $T_{\delta}$ (= 2.5 K). As shown in
Figs. \ref{fig:three}(b) and (c), the local minimum in $\delta$ at
$T_{min}$ is observed for 10 $\leq H \leq$ 35 kOe, and the drastic
decrease in $\delta$ below $T_{\delta}$ is seen for 15 $\leq H
\leq$ 35 kOe. The value of $T_{min}$ decreases with increasing
$H$, while the value of $T_{\delta}$ increases with increasing
$H$: $T_{min}$ = 15 K at $H$ = 10 kOe, 10.2 K at $H$ = 35 kOe, and
$T_{\delta}$ = 2.2 K at $H$ = 15 kOe, 4.2 K at $H$ = 35 kOe. The
$H$ dependence of $T_{min}$ and $T_{\delta}$ will be discussed in
\S \ref{disD}. For $H \geq$ 38 kOe, $\delta$ becomes negative
below 30 K and decreases with decreasing $T$. No anomaly in
$\delta$ is observed. We note that the positive local maximum of
$\delta$ around 23.7 K observed for 12.5 $\leq H \leq$ 25 kOe is
related to the occurrence of the Griffiths phase (see \S
\ref{disE}). Figures \ref{fig:fournew} (a) and (b) show the $T$
dependence of $\chi_{ZFC}$, $\chi_{FC}$, and $\delta$ at $H = 20$
and 30 kOe around $T_{\delta}$.  The susceptibility $\chi_{ZFC}$
at $H = 30$ kOe drastically increases with decreasing $T$ below
$T_{\delta}$, leading to the decrease of $\delta$.

\begin{figure*}
\begin{center}
\includegraphics[width=14.0cm]{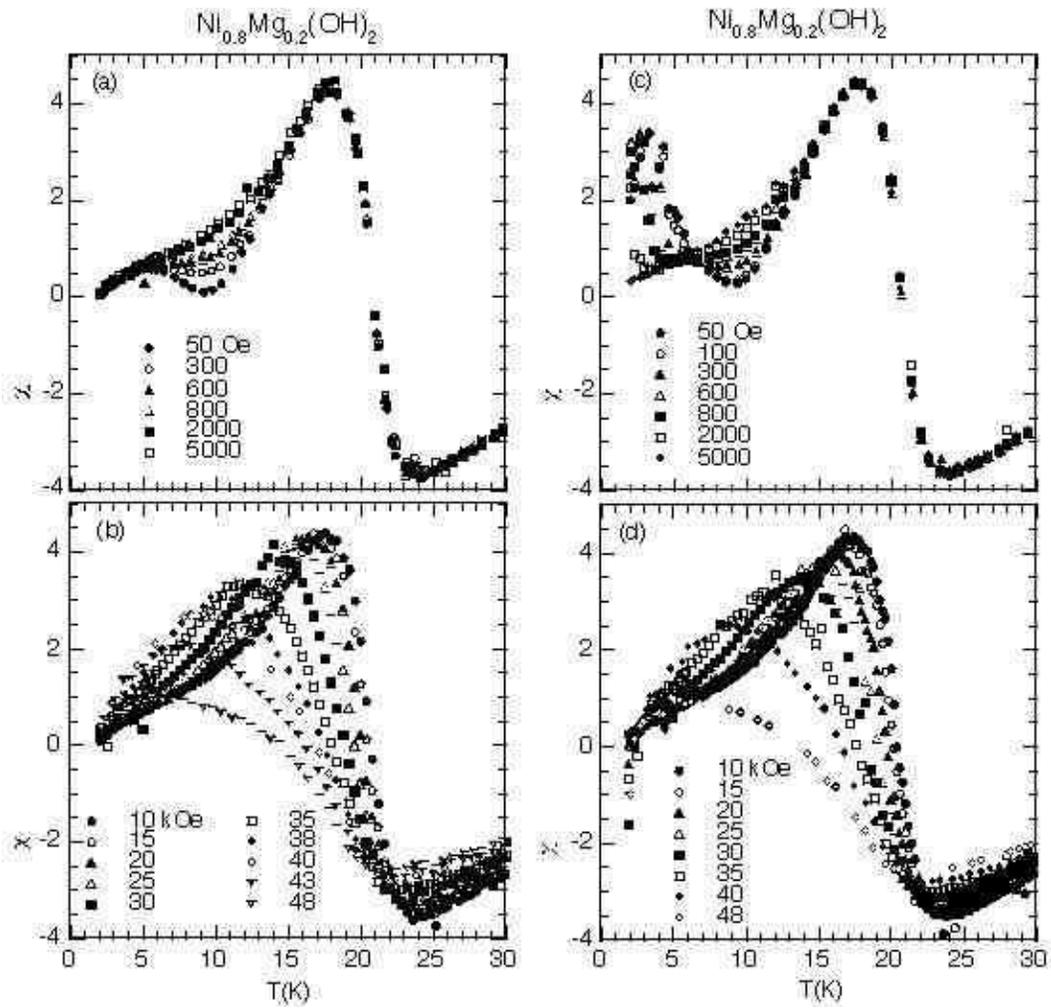}%
\end{center}
\caption{(a) and (b) $T$ dependence of d$\chi_{FC}$/d$T$
for at various $H$. (c) and (d) $T$ dependence of
d$\chi_{ZFC}$/d$T$ for at various $H$.}
\label{fig:four}
\end{figure*}

Figures \ref{fig:four}(a) and (b) show the $T$ dependence of
d$\chi_{FC}$/d$T$ for each $H$, which is calculated based on the data of
$\chi_{FC}$ vs $T$. For $H$ = 50 Oe, d$\chi_{FC}$/d$T$ shows two peaks at
5.2 and 17.9 K. The peak at low $T$ disappears above 800 Oe. The peak at
high $T$ is well defined for $H \leq$ 30 kOe and is increasingly rounded,
partly because of an increasing temperature interval where the critical
slowing-down effect predominates. Note that the temperature at which
d$\chi_{FC}$/d$T$ is equal to zero coincides with the peak temperature of
$\chi_{FC}$. Figures \ref{fig:four}(c) and (d) show the $T$ dependence of
d$\chi_{ZFC}$/d$T$ for each $H$, which is calculated based on the data of
$\chi_{ZFC}$ vs $T$. For $H$ = 50 Oe, d$\chi_{ZFC}$/d$T$ shows two peaks
at 3.4 and 17.80 K. The peak at low $T$ shifts to the low-$T$ side with
increasing $H$. The peak at high $T$ is increasingly rounded above 35 kOe.
Similar behavior is observed in Fe$_{0.46}$Zn$_{0.54}$F$_{2}$
\cite{Lederman1993}, where d$\chi_{ZFC}$/d$T$ has a sharp peak at
$T_{N}(H)$ for $H \leq$ 23 kOe and a rounded peak for $H \geq$ 40 kOe. The
$H$ dependence of the peak temperature for
Ni$_{0.8}$Mg$_{0.2}$(OH)$_{2}$ is well described by
$T_{p}(H)=T_{p}(H=0)-AH^{y}$, with $T_{p}(H=0)$ = 17.74 $\pm$ 0.02 K, $A$ =
(1.92 $\pm$ 0.87) $\times$ 10$^{-8}$, and $y$ = 1.83 $\pm$ 0.09.

\subsection{\label{resultB}$\chi^{\prime}(T,H)$}

\begin{figure}
\begin{center}
\includegraphics[width=7.5cm]{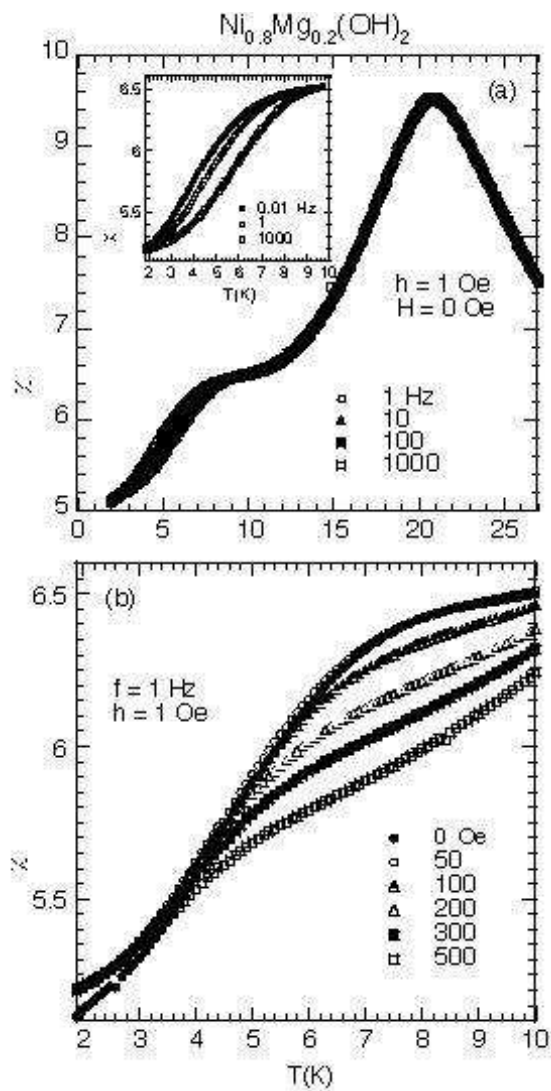}%
\end{center}
\caption{(a) $T$ dependence of $\chi^{\prime}$
at various $f$. $h$ = 1 Oe. $H$ = 0. (b) $T$ dependence of
$\chi^{\prime}$ at various $H$. $f$ = 1 Hz.
$h$ = 1 Oe.}
\label{fig:five}
\end{figure}

\begin{figure}
\begin{center}
\includegraphics[width=8.0cm]{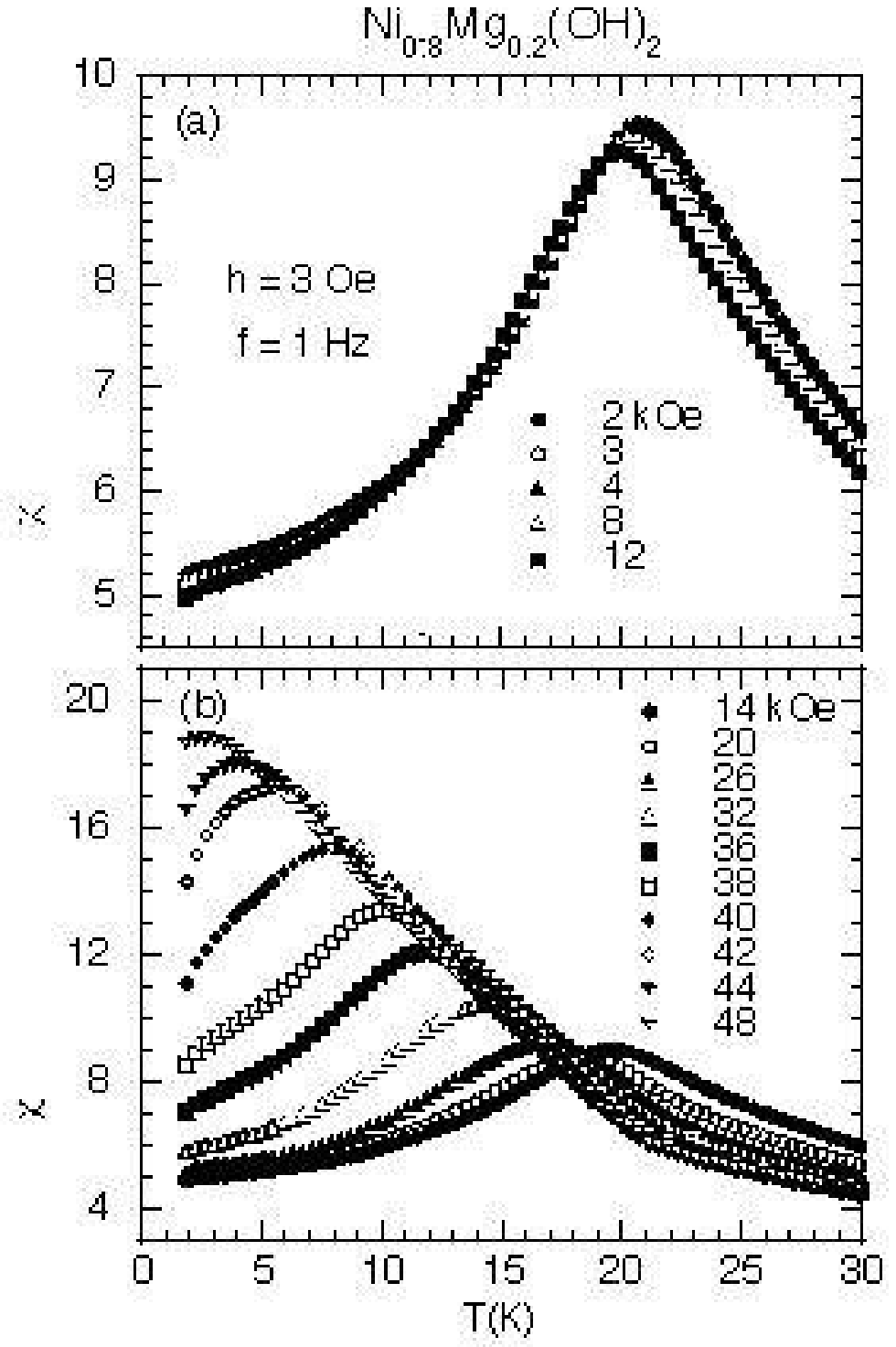}%
\end{center}
\caption{(a) and (b) $T$ dependence of $\chi^{\prime}$
at various $H$. $f$ = 1 Hz. $h$ = 3 Oe.}
\label{fig:six}
\end{figure}

Figure \ref{fig:five}(a) shows the $T$ dependence of the
dispersion $\chi^{\prime}$ at various $f$, where $H$ = 0 and $h$ =
1 Oe.  The dispersion $\chi^{\prime}$ has a broad shoulder around
5 - 7 K and a sharp peak at 20.7 K. The shape of the shoulder is
strongly dependent on $f$, while the shape of the peak is
independent of $f$.  Figure \ref{fig:five}(b) shows the $T$
dependence of $\chi^{\prime}$ at $H$ = 0 and at $H$ (= 50 - 500
Oe), where $f$ = 1 Hz and $h$ = 1 Oe.  The shoulder tends to
disappear above 500 Oe.  Figures \ref{fig:six}(a) and (b) show the
$T$ dependence of $\chi^{\prime}$ at higher $H$, where $h$ = 3 Oe
and $f$ = 1 Hz.  The shoulder completely disappears at $H$ = 2
kOe.  For $H \leq$ 14 kOe the peak slightly shifts to the low-$T$
side with increasing $H$.  For $H \geq$ 20 kOe the peak
drastically shifts to the low-$T$ side.  As will be discussed in
\S \ref{disA}, it is reasonable to define the peak temperature of
$\chi^{\prime}$ vs $T$ as $T_{N}(H)$.  The relation between
$T_{N}(H)$ and $H$ leads to the $H$-$T$ phase diagram.  We find
that the curvature of $H$ vs $T$ changes from concave for $H >$ 38
kOe to convex for $H <$ 38 kOe.  Note that there are two kinds of
field-induced transitions for AF systems: the spin-flop transition
for $H_{E}^{\prime} > H_{A}$ and the metamagnetic transition for
$H_{E}^{\prime} < H_{A}$.  Here $H_{E}^{\prime}$ is the
interplanar exchange field and $H_{A}$ is the anisotropy field.
The spin-flop transition occurs between the AF and spin-flop (SF)
phases at
$H_{SF}=[H_{A}(2H_{E}^{\prime}-H_{A})]^{1/2}$.\cite{Suzuki2002}.
The magnetization saturates at $H_{s}=2H_{E}^{\prime}-H_{A}
\approx 2H_{E}^{\prime}$.  In contrast, the metamagnetic
transition occurs between the AF and ferromagnetic (FM) phases at
$H=H_{E}^{\prime}$.  It is considered that our system with $c$ =
0.8 undergoes the spin-flop transition rather than the
metamagnetic transition.  When $H_{s} \approx$ 55 kOe for $c$ = 1
\cite{Enoki1979} and $H_{SF}$ is assumed to be 38 kOe, we can
estimate $H_{E}^{\prime}$ = 40.6 kOe and $H_{A}$ = 26.3 kOe,
satisfying the condition for the spin-flop transition.  The value
of $H_{E}^{\prime}$ thus obtained is larger than that
($H_{E}^{\prime} \approx 22.3$ kOe) reported by Enoki et al.  for
$c$ = 1 \cite{Enoki1975a,Enoki1979}.  The AF, the paramagnetic
(PM) phase, and a possible SF phase merge at a critical point
$P_{0}$ ($H_{0} = 38$ kOe, $T_{0} = 10$ K).  A line connecting
between the critical point $P_{0}$ and $T$ = 20.8 K at $H$ = 0 is
a critical line $H_{c}$ of the second order (see the definition of
$H_{c}$ and $P_{0}$ in \S \ref{disB}). This critical line is well
described by
\begin{equation}
T_{N}(H) = T_{N}(H=0) - AH^{y},
\label{eq:two}
\end{equation}
with $T_{N}(c=0.8,H=0)$ = 20.76 $\pm$ 0.10 K, $A$ = (5.83 $\pm$ 1.97)
$\times$ 10$^{-9}$, and $y$ = 2.00 $\pm$ 0.03. The index $y$ is very close
to the prediction from the molecular field theory.

\subsection{\label{resultC}$\chi^{\prime \prime}(T,H)$}

\begin{figure}
\begin{center}
\includegraphics[width=7.5cm]{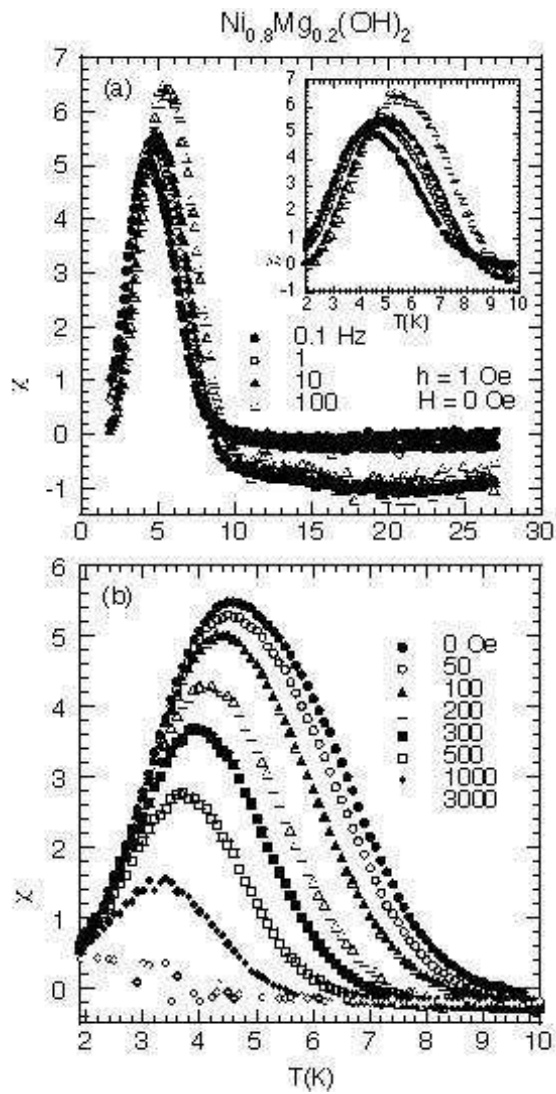}%
\end{center}
\caption{(a) $T$ dependence of $\chi^{\prime \prime}$
at various $f$. $h$ = 1 Oe. $H$ = 0. (b) $T$ dependence of
$\chi^{\prime \prime}$ at various $H$. $f$ = 1 Hz. $h$ = 1
Oe \cite{Suzuki2000}.}
\label{fig:seven}
\end{figure}

Figure \ref{fig:seven}(a) shows the $T$ dependence of the absorption
$\chi^{\prime \prime}$ at various $f$, where $H$ = 0 and $h$
= 1 Oe. No anomaly is observed around $T_{N}$ \cite{Suzuki2000}.
A peak observed around
4.55 K is strongly dependent on $f$. This peak shifts to the high-$T$ side
with increasing $f$. This peak is related to the RSG ordering. Figure
\ref{fig:seven}(b) shows the $T$ dependence of $\chi^{\prime \prime}$ for
at $H$ = 0 and at $H$ ($H$ = 50 Oe - 3 kOe), where $f$ = 1 Hz and
$h$ = 1 Oe. The peak shifts to the low-$T$ side with increasing $H$ : 2.68
K at $H$ = 2 kOe and 2.3 K at $H$ = 3 kOe. It tends to disappear at $H$ =
3 kOe. In the inset of Fig. \ref{fig:three}(a), for comparison, we show
the $H$ dependence of $T_{RSG}$, which is defined as the peak temperature
for the data of $\chi^{\prime \prime}$ vs $T$. Note that $T_{RSG}$ from
$\delta$ is a little higher than that from $\chi^{\prime \prime}$ at the
same $H$.

\begin{figure}
\begin{center}
\includegraphics[width=7.5cm]{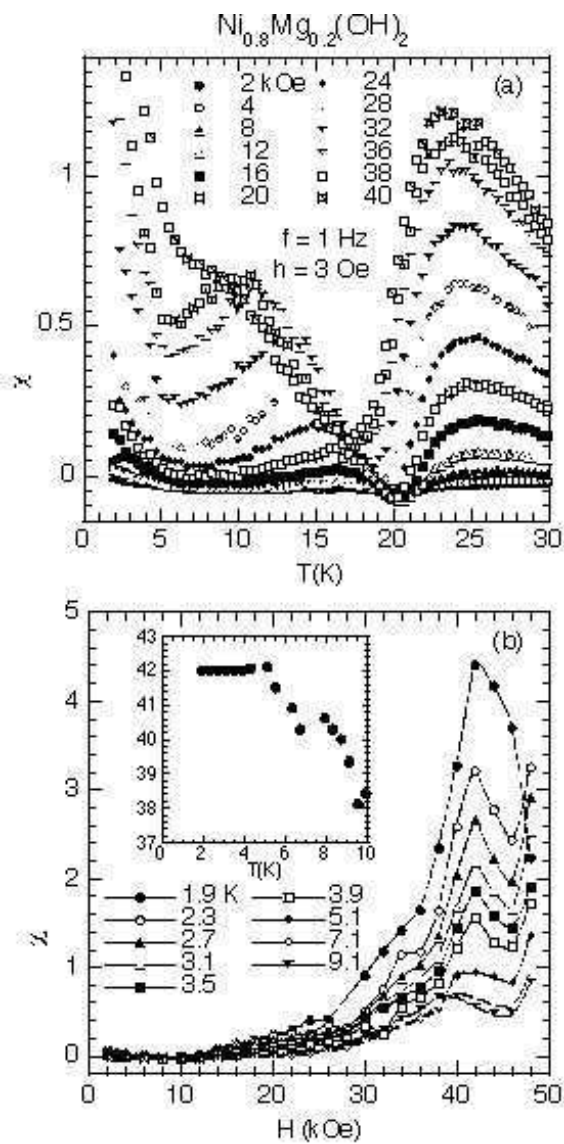}%
\end{center}
\caption{(a) $T$ dependence of $\chi^{\prime \prime}$
at various $H$. (b) $H$ dependence of $\chi^{\prime \prime}$
at various $T$. $f$ = 1 Hz. $h$ = 3 Oe. The inset shows the
plot of the peak field vs $T$.}
\label{fig:eight}
\end{figure}

Figure \ref{fig:eight}(a) shows the $T$ dependence of the absorption
$\chi^{\prime \prime}$ at higher $H$, where $h$ = 3 Oe and $f$ = 1 Hz. A
strong divergence of $\chi^{\prime \prime}$ for $H \geq$ 40 kOe is observed
as $T$ approaches zero. This is in contrast to a weak increase of
$\chi^{\prime \prime}$ for $H \leq$ 30 kOe with decreasing $T$. Figure
\ref{fig:eight}(b) shows the $H$ dependence of $\chi^{\prime \prime}$ at
various low $T$. The absorption $\chi^{\prime \prime}$ drastically
increases with increasing $H$, showing a sharp peak around 42 kOe. This
result strongly suggests the transition between the RSG and a SG phases, as
is predicted by Takayama \cite{Takayama1988} for the AF-RSG systems (see
Fig. \ref{fig:one}).

As shown in Fig. \ref{fig:eight}(a) $\chi^{\prime \prime}$ for 12 $\leq H
\leq$ 40 kOe exhibits a rather broad peak around 8.2 - 16.5 K, which is due
to the PM-AF transition. The peak shifts to the low-$T$ side with
increasing $H$: 16.5 K at $H$ = 12 kOe and 8.2 K at 40 kOe. This peak
temperature is a little lower than the peak temperature of $\chi^{\prime}$
or $T_{N}(H)$ at the same $H$. We also note another broad peak in
$\chi^{\prime \prime}$ at $T$ above $T_{N}(H)$ for $H \geq$ 10 kOe. The
dispersion $\chi^{\prime}$ does not show any anomaly in the same region of
the $H$-$T$ plane. The broad peak in $\chi^{\prime \prime}$ shifts to the
low-$T$ side with increasing $H$: 26.0 K at $H$ = 10 kOe and 22.3 K at $H$
= 48 kOe. This peak is not due to the ferromagnetic short-range
fluctuations occurring in the same layer, since the peak temperature
decreases with increasing $H$. This peak is related to the Griffiths
phase. The absorption $\chi^{\prime \prime}$ at $H$ = 2 kOe shows a local
minimum at 21.1 K. This minimum survives up to 48 kOe: it shifts to the
low-$T$ side with increasing $H$. The $H$ dependence of local-minimum
temperature is almost the same as that of the peak temperatures of
$\chi_{ZFC}$ and $\chi_{FC}$.

\begin{figure}
\begin{center}
\includegraphics[width=7.5cm]{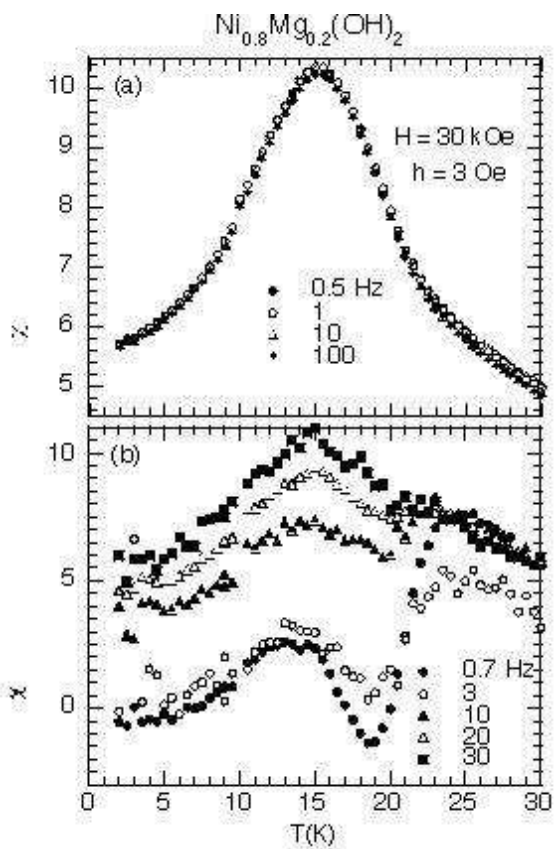}%
\end{center}
\caption{$T$ dependence of (a) $\chi^{\prime}$ and (b)
$\chi^{\prime \prime}$ at various $f$ (0.5 $\leq f \leq$
100 Hz). $H$ = 30 kOe. $h$ = 3 Oe.}
\label{fig:nine}
\end{figure}

\begin{figure}
\begin{center}
\includegraphics[width=8.0cm]{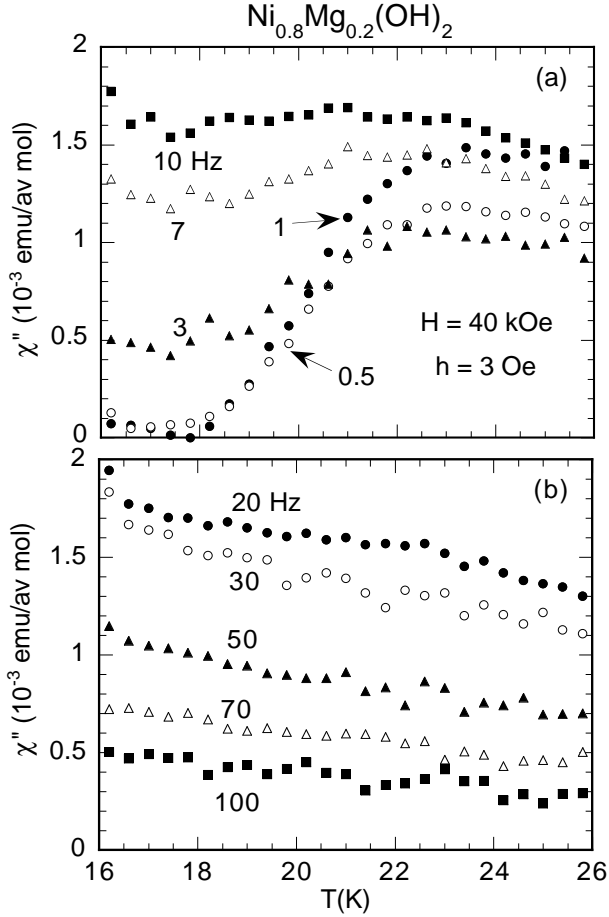}%
\end{center}
\caption{(a) and (b) $T$ dependence of $\chi^{\prime
\prime}$ at various $f$ (0.5 $\leq f \leq$ 100 Hz). $H$ = 40
kOe. $h$ = 3 Oe.}
\label{fig:ten}
\end{figure}

In order to investigate the nature of the Griffiths phase we have measured
the $f$ and $T$ dependence of $\chi^{\prime}$ and $\chi^{\prime
\prime}$ at $H$ = 30 and 40 kOe. Figures \ref{fig:nine}(a) and (b) show
the $T$ dependence of $\chi^{\prime}$ and $\chi^{\prime \prime}$ at $H$ =
30 kOe, respectively, where $h$ = 3 Oe and $f$ is varied between 0.5 and
100 Hz. The dispersion $\chi^{\prime}$ has a single peak at 15.2 K, which
is independent of $f$. In contrast, $\chi^{\prime \prime}$ shows two peaks
at 12.95 K and $T_{G} \approx$ 25.1 K at $f$ = 0.5 Hz. The peak at 25.1 K
slightly shifts to the low-$T$ side with increasing $f$ ($T_{G}$ = 22.1 K
for $f$ = 20 Hz) and disappears for $f \geq$ 30 Hz. Figures
\ref{fig:ten}(a) and (b) show the $T$ dependence of $\chi^{\prime \prime}$
at $H$ = 40 kOe, where $h$ = 3 Oe and $f$ is varied between 0.5 and 100 Hz.
The absorption $\chi^{\prime \prime}$ at $f$ = 0.5 Hz shows a peak around
23.5 K. This peak shifts to the low-$T$ side with increasing $f$ for 0.5
$\leq f \leq$ 3 Hz and disappears above 20 Hz. It is predicted that the
clusters in the Griffiths phase are nearly ordered and relax very slowly.
The $T$ dependence of the average relaxation time ($\tau_{G}$) will be
discussed in \S \ref{disE}.

\section{\label{dis}Discussion}
\subsection{\label{disA}Definition of $T_{N}(H)$}

\begin{figure}
\begin{center}
\includegraphics[width=7.5cm]{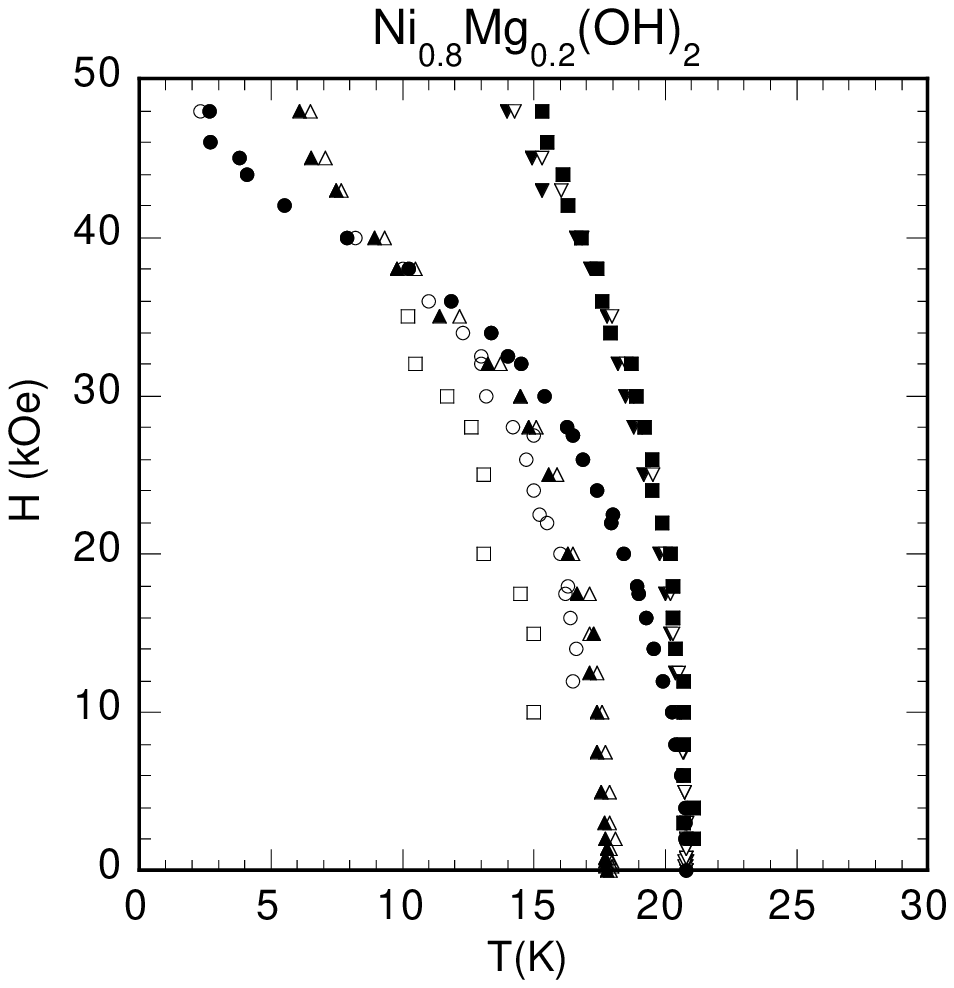}%
\end{center}
\caption{Definition of $T_{N}(H)$. The
following characteristic temperatures are plotted as a function of $H$: the
peak temperatures of $\chi^{\prime}$ ({\Large $\bullet$}), $\chi^{\prime \prime}$
({\Large $\circ$}), d$\chi_{ZFC}$/d$T$ ($\blacktriangle$), d$\chi_{FC}$/d$T$
($\triangle$), $\chi_{ZFC}$ ($\blacktriangledown$), and $\chi_{FC}$
($\triangledown$), and local-minimum temperatures of $\chi^{\prime \prime}$
($\blacksquare$) and $\delta$ ($\square$).}
\label{fig:thirteen}
\end{figure}

Before we discuss the $H$-$T$ phase diagram, it is instructive to explain
how to determine the N\'{e}el temperature $T_{N}(H)$. In Fig.
\ref{fig:thirteen} we make a plot of the following characteristic
temperatures which may be related to $T_{N}(H)$, as a function of $H$ in
the $H$-$T$ plane: the peak temperatures of $\chi_{FC}$ vs $T$,
$\chi_{ZFC}$ vs $T$, d$\chi_{FC}$/d$T$ vs $T$, d$\chi_{ZFC}$/d$T$ vs $T$,
$\chi^{\prime}$ vs $T$, and $\chi^{\prime \prime}$ vs $T$, and
local-minimum temperatures of $\chi^{\prime \prime}$ vs $T$ and $\delta$ vs
$T$. These characteristic temperatures can be classed into four groups,
depending on $H$ and $T$: (i) the peak temperatures of $\chi_{FC}$,
$\chi_{ZFC}$, and local-minimum temperature of $\chi^{\prime \prime}$, (ii)
the peak temperature of $\chi^{\prime}$, (iii) the peak temperatures of
d$\chi_{FC}$/d$T$, d$\chi_{ZFC}$/d$T$, and $\chi^{\prime \prime}$, and (iv)
the local-minimum temperature of $\delta$. In an usual Ising
antiferromagnet, the $T$-derivative of the DC susceptibility
$\chi_{\parallel}$ along the easy axis, (d$\chi_{\parallel}$/d$T$),
diverges at $T_{N}(H)$. The peak temperature of $\chi_{\parallel}$ vs $T$
is higher than $T_{N}(H)$. The magnetic heat capacity $C_{mag}$ as well as
$\chi^{\prime}$ and $\chi^{\prime \prime}$ also show a peak at $T_{N}(H)$.
Because of the Maxwell relation [($\partial$$S$/$\partial$$H$)$_{T}$ =
($\partial$$M$/$\partial$$T$)$_{H}$], the $H$-derivative of $C_{mag}$ is
related to d$\chi_{\parallel}$/d$T$. Thus the first and fourth groups are
not directly related to $T_{N}(H)$. The second and third groups of our
system may correspond to the critical lines denoted by the lines $H_{c}$
and $H_{1}$ (see Fig. \ref{fig:fourteen}). Similar behavior is
observed in a 3D Ising antiferromagnet
FeBr$_{2}$ \cite{Azevedo1995,Katori1996,Katsumata1997,Petracic1998,Bine2000}.
In FeBr$_{2}$, the data of $C_{mag}$ vs $T$, $\chi^{\prime}$ vs $T$, and
$\chi^{\prime \prime}$ vs $T$ show peaks at $T_{N}(H)$ along the line
$H_{c}$, and the data of $C_{mag}$ vs $T$ and $\chi^{\prime \prime}$ vs $T$
show peaks at a critical temperature $T_{1}(H)$ ($< T_{N}(H)$) along the
line $H_{1}$.

\subsection{\label{disB}$H$-$T$ phase diagram}

\begin{figure}
\begin{center}
\includegraphics[width=7.5cm]{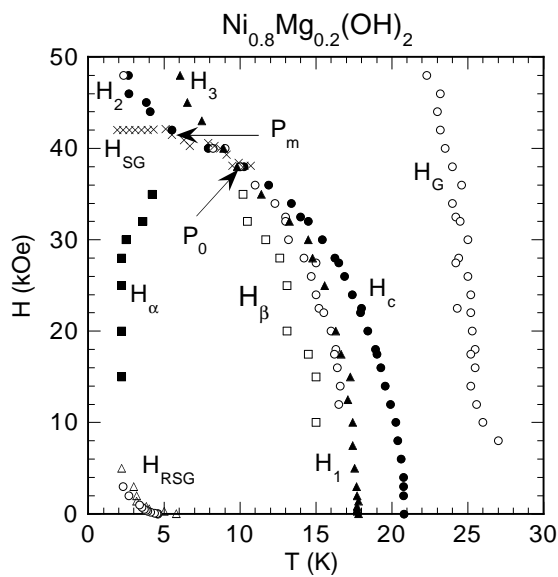}%
\end{center}
\caption{$H$-$T$ phase diagram: peak
temperatures of $\chi^{\prime}$ ({\Large $\bullet$}), $\chi^{\prime \prime}$
({\Large $\circ$}), d$\chi_{ZFC}$/d$T$ ($\blacktriangle$), the
characteristic temperatures $T_{\delta}$ ($\blacksquare$), $T_{min}$
($\square$), and $T_{RSG}$ ($\triangle$) for $\delta$. The definition of
lines $H_{c}$, $H_{1}$, $H_{2}$, $H_{3}$, $H_{\alpha}$, $H_{\beta}$,
$H_{RSG}$, $H_{G}$, the points $P_{m}$ and $P_{0}$ is given in the text.}
\label{fig:fourteen}
\end{figure}

Figure \ref{fig:fourteen} shows the $H$-$T$ phase diagram of
Ni$_{c}$Mg$_{1-c}$(OH)$_{2}$ with $c$ = 0.8. It consists of the lines
denoted by $H_{c}$, $H_{1}$, $H_{2}$, $H_{3}$, $H_{\alpha}$, $H_{\beta}$,
$H_{RSG}$, and $H_{G}$, and two points denoted by a multicritical point
$P_{m}$ ($T_{m}$ = 5.6 K, $H_{m}$ = 42 kOe) and a critical point $P_{0}$
($T_{0}$ = 10 K, $H_{0}$= 38 kOe). The
features of the $H$-$T$ diagram are summarized as follows. (i) A spin-flop
transition occurs around the critical point $P_{0}$. The line $H_{c}$
connecting between $P_{0}$ and $T_{N}(c=0.8,H=0)$ (= 20.7 K), is of the
second order. (ii) The region enclosed by the
extension of the lines $H_{c}$, $H_{2}$, and $H_{3}$ may be a spin-flop (SF)
phase. (iii) The line $H_{1}$ connects between the critical point $P_{0}$
and a characteristic temperature (= 17.74 K) at $H$ = 0. The nature of
intermediate phase between the lines $H_{1}$ and $H_{c}$ is not clear at
present. Note that similar intermediate phase is observed in FeBr$_{2}$,
where the transverse component of AF order parameter is enhanced on
crossing the line $H_{1}$ \cite{Bine2000}. (iv) The line $H_{RSG}$ is the
boundary between the RSG and AF phases. This line corresponds to an
Almeida-Thouless (AT) line between SG and PM phases \cite{Almeida1978}.
(v) The line
$H_{SG}$ is the boundary between the SG and RSG phases (see \S
\ref{disC}). The region enclosed by the lines $H_{SG}$ and $H_{2}$ is the
SG phase. (vi) The extensions of the lines $H_{\alpha}$ and $H_{\beta}$
may merge at the multicritical point $P_{m}$. The nature of the lines
$H_{\alpha}$ and $H_{\beta}$ will be discussed in \S \ref{disD}. (vii)
The line $H_{G}$ is in the PM phase and is related to the
Griffiths phase (see \S \ref{disE}). The characteristic temperatures
along the line $H_{G}$ are lower than $T_{N}(c=1,H=0)$ (= 26.4 K) of the
pure system. It slightly decreases with increasing $H$, reflecting the
antiferromagnetic nature of the system. The line $H_{G}$ is essentially
different from a non-critical line for ferromagnetic short-range order
where a characteristic temperature increases with increasing $H$ in the
PM region.

\subsection{\label{disC}Field-induced spin glass phase}
We compare the $H$-$T$ phase diagram for Ni$_{0.8}$Mg$_{0.2}$(OH)$_{2}$
(Fig. \ref{fig:fourteen}) with the predicted one shown in Figs.
\ref{fig:one}(a) and (b). The lines $H_{2}$ and $H_{SG}$ meet the
extension of the line $H_{c}$ at the multicritical point $P_{m}$. The line
$H_{SG}$ is the boundary between the RSG and SG phases. There is only a SG
state in the SG phase, while the SG phase coexists with the AF phase in the
RSG phase. The line $H_{RSG}$, which is the boundary between the RSG and
AF phase, disappears above 5 kOe. If the line $H_{\alpha}$ corresponds to
the boundary between the RSG and AF phases, our $H$-$T$ diagram
is in good agreement with the prediction shown in Fig.
\ref{fig:one}(a): the AF-RSG transition temperature increases with
increasing $H$. The extrapolation of the line $H_{\alpha}$ to the
higher-$T$ side seems to intersect the multicritical point $P_{m}$. The
nature of the line $H_{\alpha}$ will be discussed in \S \ref{disD}.
Here we note that only $\chi^{\prime \prime}$ exhibits an anomaly when
crossing the line $H_{SG}$ at fixed $T$.

The $H$-$T$ phase diagram for Ni$_{0.8}$Mg$_{0.2}$(OH)$_{2}$
is a little different from that of Fe$_{0.552}$Mg$_{0.448}$Cl$_{2}$
which undergoes a metamagnetic transition around a multicritical point. In
Fe$_{0.552}$Mg$_{0.448}$Cl$_{2}$ the line $H_{RSG}$ intersects the
line $H_{c}$ at the multicritical point $P_{m}$ \cite{Gelard1983}. There are
two features on the data of $T_{RSG}(H)$ vs $H$. The critical temperature
$T_{RSG}(H)$ at low $H$ increases with increasing $H$, enlarging the domain
of existence of RSG phase. The temperature $T_{RSG}$ starts to decrease at
higher $H$ before it meets the multicritical point. The former feature is
in good agreement with the prediction from the molecular field theory, but
the latter feature is rather different from the prediction.

Along the line $H_{2}$ a striking reversal of the curvature of the line
$H_{2}$ is observed as $H$ is increased above 42 kOe. Similar behavior is
observed for Fe$_{c}$Mn$_{1-c}$TiO$_{3}$ with $c$ = 0.60 and
0.65 \cite{Yoshizawa1994,Katori1992}. The line $H_{2}$ is considered to be
associated with the Almeida-Thouless (AT) replica symmetry-breaking line.
In fact the curvature of the line $H_{2}$ seems to be similar to that of
the line $H_{RSG}$. The reversal of curvature is understood within the
framework of the mean field theory \cite{Takayama1988}. The origin of the SG
phase for $c$ = 0.8 is considered to be as follows \cite{Suzuki2000}. The dilution by
nonmagnetic Mg introduces the competition between the intraplanar nearest
neighbor (N.N.) FM interactions, the intraplanar next-nearest-neighbor
(N.N.N.) AF interactions, and the interplanar N.N. AF interactions, which
leads to the spin frustration effect.

\subsection{\label{disD}Glassy phase}
The nature of the irreversibility of susceptibility ($\delta$) is discussed
in relation to glassy phases in SG, RSG, Griffiths phases, and RF effect.
Our results of $\delta$ (see Figs. \ref{fig:three} and
\ref{fig:fournew}) are
summarized as follows. (i) For $H <$ 10 kOe, $\delta$ is always positive
for 1.9 $\leq T \leq$ 30 K. (ii) For $H \leq$ 3 kOe, $\delta$ increases in
two steps with decreasing $T$: a weak change at $T_{RF}$ and a strong
change at $T_{RSG}$. (iii) For $H >$ 12.5 kOe, $\delta$ shows a negative
local minimum at $T_{min}$. (iv) For 10 $\leq H \leq$ 25 kOe, $\delta$
shows a local maximum around the Griffiths temperature $T_{G}(H)$ above
$T_{N}(H)$. (v) At 17.5 $\leq H \leq$ 35 kOe, $\delta$ shows a drastic
decrease with decreasing $T$ below $T_{\delta}$, as well as a negative
local minimum at $T_{min}$. (vi) For $H \geq$ 38 kOe, $\delta$ becomes
negative below 30 K and decreases with decreasing $T$. In the $H$-$T$
diagram of Fig. \ref{fig:fourteen}, we make a plot of $T_{\delta}$,
$T_{min}$, and $T_{RSG}$ for each $H$, which forms the line $H_{\alpha}$
for $T_{\delta}$, the line $H_{\beta}$ for $T_{min}$, and the line
$H_{RSG}$ for $T_{RSG}$. The extension of the lines $H_{\alpha}$ and
$H_{\beta}$ seems to join the multicritical point $P_{m}$. The transition
temperature $T_{RSG}(H)$ defined as the onset temperature of strong
irreversibility in $\delta$ is a little higher than the peak temperature of
$\chi^{\prime \prime}$ (see the inset of Fig. \ref{fig:three}(a)). No RSG
transition is observed above 5 kOe.

We note that the two step-like changes in $\delta$ is also
observed in the 3D Ising spin glass Fe$_{c}$Mn$_{1-c}$TiO$_{3}$. This system
undergoes the P-AF and AF-RSG transitions at $H$ = 0. The transition
temperatures $T_{RSG}(H)$ and $T_{N}(H)$ are defined as the peak
temperatures of magnetic diffuse neutron scattering, where the inverse spin
correlation length has local minima \cite{Yoshizawa1994}.
A series of magnetization
measurements \cite{Katori1992} on Fe$_{c}$Mn$_{1-c}$TiO$_{3}$ (0 $< c <$
0.41) indicates that $T_{RSG}(H)$ is not the onset temperature below which
$\delta$ become negative. The difference $\delta$ exhibits a crossover
from a weak irreversibility just below $T_{N}(H)$ to a strong
irreversibility at $T_{RSG}(H)$. The weak irreversibility is related to
the RF effect and the strong irreversibility is related to the RSG
transition. In fact, Yoshizawa et al. \cite{Yoshizawa1994} have shown from
the magnetic neutron scattering of Fe$_{0.6}$Mn$_{0.4}$TiO$_{3}$
that the antiferromagnetic intensity and diffuse scattering suffer from
the RF effect in the presence of $H$, in spite of the fact that
Fe$_{0.6}$Mn$_{0.4}$TiO$_{3}$ is not a diluted Ising
antiferromagnet. The neutron scattering profile as well as the temperature
dependence of the intensities exhibit pronounced history dependence between
the FC and ZFC states below $T_{N}(H)$. The ZFC measurement shows a
sharp transition to a long-range ordered AF state, while the FC measurement
exhibits broadened transitions characteristic of domains (short range)
being frozen below $T_{N}(H)$. The FC procedure causes the system to lose
equilibrium and enter a metastable, frozen, domain state without long-range
order.

How can we understand the anomalous behavior in $\delta$ near the lines
$H_{\beta}$? The line $H_{\beta}$ is
qualitatively explained as follows. Numerical
calculations \cite{Soukoulis1983,Ro1985,Soukoulis1985,Grest1986,Almeida1987}
based on the molecular field theory for a 3D diluted Ising antiferromagnet
suggest that a region of AF long-range order and a region of glassy state
exist in the $H$-$T$ plane. In the region of AF long-range order, the ZFC
state has a lower free energy and in the region of glassy state the FC
state does. de Almeida and Bruinsma \cite{Almeida1987} have shown that the
region of glassy state is truly separate phase. The glassy phase remains
restricted to large $H$, where the strength of random field varies with
$H$. The detail of the phase boundary between these two phases has been
obtained by Ro et al. \cite{Ro1985} for the 3D diluted AF system with $c$ =
0.7. The line $H_{\beta}$ qualitatively coincides with this
phase boundary. The origin of this line may
be due to inhomogeneous internal field, which is given by $H_{int}$ =
$H-NM$, where $N$ is the demagnetization factor. This internal field
couples with the AF order parameter. If this field is strongly
inhomogeneous, it may play the same role as the RF field. The line
$H_{\alpha}$ is not predicted from numerical
calculations \cite{Soukoulis1983,Ro1985,Soukoulis1985,Grest1986,Almeida1987}.
The line $H_{\alpha}$ may correspond to the
boundary between the RSG and AF phases.

How does the random field effect occur in our powdered sample which are
assumed to be formed of nanoparticles?  The local magnetic field for
spin inside each nanoparticle is described by $H_{i}$ $\cos \theta$, where $\theta$ (0
$\leq \theta \leq 2\pi$) is the angle between the local magnetic field {\bf
H$_{i}$} and the easy spin axis of each particle.  The mean value of the
effective field is $[H_{i}]_{av}$$<$$\cos \theta$$>$ = 0, while the
variance of the effective field is $[H_{i}^{2}]_{av}$$<$$\cos^{2}
\theta$$>$ $= [H_{i}^{2}]_{av}/2 = (H_{R})^{2}/2$, where $<$$\cos^{2}
\theta$$> = 1/2$ and $H_{R}$ is the random field of the single crystal and
is defined as $H_{R}^{2} = [H_{i}^{2}]_{av}$ \cite{Mydosh}.  Thus the random field for
the powdered system is $H_{R}/\sqrt{2}$.  In this sense, the random field
effect is still effective for the powdered system.

\subsection{\label{disE}Griffiths phase}
Experimentally, the existence of the unfrustrated clusters of spins
is confirmed from the maxima of $\delta$ vs $T$ and
$\chi^{\prime \prime}$ vs $T$ in the PM phase above $T_{N}(c,H)$.
Such clusters may be in nearly ordered states, and fluctuates with large
relaxation times. This phenomenon is understood as the occurrence of the
Griffiths phase. Here we briefly describe the picture of the cluster
dynamics in the Griffiths phase. In a dilute Ising ferromagnet with
concentration $c$, a cluster is defined as a compact region in which spins
are connected with each other by bonds. Such clusters become quasi-ordered
at temperatures below $T_{G}$ and flips collectively with anomalous long
characteristic times. The consequence of such processes is slow dynamics
of the spin auto-correlation function $q(t)$, whose asymptotic form at a
large time $t$ is given by $q(t) \approx$ exp$[-\lambda($ln
$t)^{d/(d-1)}]$ \cite{Komori1997,Bray1987,Bray1988}, where $d$ is the
dimension of the system and $\lambda$ is a constant. This decay form is
called the enhanced power law. Griffiths \cite{Griffiths1969} has shown
that for a randomly site-diluted Ising ferromagnet with magnetic
concentration $c$, the magnetization fails to be an analytical function of
$H$ for any temperature within the so-called Griffiths phase $T_{c}(c,H)
\leq T_{G}(H) \leq T_{c}(c=1,H=0)$, where $T_{c}(c,H)$ is the critical
temperature at $H$ for the diluted ferromagnetic system with the
concentration $c$. In our system, Ni spins with Ising symmetry are
ferromagnetically ordered in the same Ni$_{c}$Mg$_{1-c}$ layer, where a
part of Ni$^{2+}$ spins is randomly replaced by nonmagnetic Mg$^{2+}$ ions.
Thus in the limit of weak AF interplanar interactions, our system is
similar to the site-diluted Ising ferromagnet: $T_{N}(c,H) \leq T_{G} \leq
T_{N}(c=1,H=0)$. Surprisingly the Griffiths phase is observed even for the
pure system Ni(OH)$_{2}$, where $T_{G}(c=1,H)$ is larger than
$T_{N}(c=1,H=0)$. Such a Griffiths phase is also observed in a 3D
antiferromagnet FeCl$_{2}$ \cite{Binek1994}. For FeCl$_{2}$, $T_{G}(H)$ is
between $T_{N}(c=1,H)$ and $T_{N}(c=1,H=0)$, which is different from that for
Ni(OH)$_{2}$. This behavior is explained in terms of fluctuating
distributions of demagnetizing fields.

\begin{figure}
\begin{center}
\includegraphics[width=7.5cm]{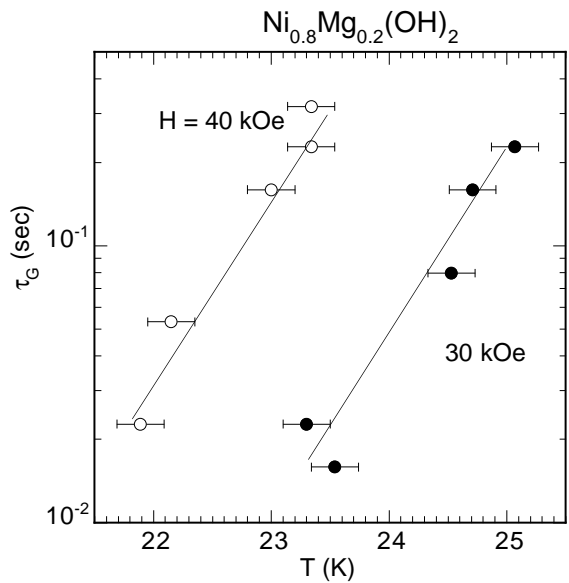}%
\end{center}
\caption{$T$ dependence of the relaxation time
$\tau_{G}$ for the Griffiths phase. $H$ = 30 and 40 kOe.
The solid lines are a guide to eyes.}
\label{fig:fifteen}
\end{figure}

What is the nature of the Griffiths phase for
Ni$_{0.8}$Mg$_{0.2}$(OH)$_{2}$? The absorption
$\chi^{\prime \prime}$ is related to the power spectrum of the
magnetization fluctuation according to the fluctuation-dissipation theorem.
As shown in Fig. \ref{fig:eight}(a), we find that $\chi^{\prime \prime}$
shows a broad peak at $T_{G}(H)$ between $T_{N}(c=0.8,H=0)$
(= 20.7 K) and $T_{N}(c=1,H=0)$ (= 26.4 K) for $H \geq$ 10 kOe. This broad
peak shifts to the low-$T$ side with increasing $H$, reflecting the nature
of antiferromagnetic fluctuations, but not ferromagnetic fluctuations.
There is no anomaly in $\chi^{\prime}$ at any $T$ above $T_{N}(H)$. As
shown in Figs. \ref{fig:nine}(b) and \ref{fig:ten}(a), the broad peak at
$T_{G}(H)$ for $H$ = 30 and 40 kOe is strongly dependent on $f$. This peak
shifts to the low-$T$ side with increasing $f$ for 0.5 $\leq f \leq$ 10 Hz
and seems to disappear for $f \geq$ 20 - 30 Hz. If $\chi^{\prime \prime}$
has a peak at $\omega \tau_{G}$ = 1 ($\omega=2\pi f$), then the
corresponding characteristic relaxation time $\tau_{G}$ can be estimated as
$\tau_{G}=(2\pi f)^{-1}$. Figure \ref{fig:fifteen} shows the $T$
dependence of $\tau_{G}$ thus obtained for $H$ = 30 and 40 kOe.
Surprisingly the relaxation time $\tau_{G}$ increases with increasing
$T$ in
the limited $T$ range. The value of $\tau_{G}$ is anomalously large at $T$
much higher than $T_{N}(H)$. Such a large $\tau_{G}$ may be attributed to
preferentially formed large spin clusters, leading to the field-induced
Griffiths phase. The reason of the increase of $\tau_{G}$ with increasing
$T$ is not clear at present.

Similar behavior is observed in Fe$_{0.47}$Zn$_{0.53}$F$_{2}$
\cite{Binek1995} and FeCl$_{2}$ \cite{Binek1994}. In
Fe$_{0.47}$Zn$_{0.53}$F$_{2}$, $\chi^{\prime}$ at $f$ = 1 Hz in
the ZFC state shows a narrow peak at $T_{N}(H)$ and a very broad main peak
at $T_{G}(H)$ [$> T_{N}(c,H)$] for $H \geq$ 30 kOe. This feature is more
pronounced as $H$ is increased. In contrast, $\chi^{\prime \prime}$ shows
a sharp peak at $T_{N}(c,H)$. It extends an asymmetric flat tail for
$T_{N}(c=0.47,H) < T < T_{N}(c=0.47,H=0)$. In FeCl$_{2}$, $\chi^{\prime
\prime}$ shows a broad peak at $T_{G}(H)$ for $H \geq$ 12 kOe, where
$T_{N}(c=1,H) < T_{G}(H) < T_{N}(c=1,H=0)$. The absorption $\chi^{\prime
\prime}$ is strongly dependent on $f$. For example, $\chi^{\prime \prime}$
at $H$ = 18 kOe has a peak at $T_{G}(H)$ at $f$ = 1 Hz. However, this peak
disappears at $f$ = 0.1 and 10 Hz. This result is in good agreement with
our result.

\section{Conclusion}
The $H$-$T$ phase diagram of Ni$_{0.8}$Mg$_{0.2}$(OH)$_{2}$
consists of the RSG, SG, and AF phases. These phases meet a multicritical
point $P_{m}$. A spin-flop transition occurs around the critical point
$P_{0}$. These results are in good agreement with the theoretical
prediction from the molecular field theory. The existence of the glassy
phase in the AF phase and the Griffiths phase in the paramagnetic phase is
also confirmed. Further studies will be required to understand the
nature of the SG phase, glassy phase and Griffiths phase.

\section*{Acknowledgment}
The authors would like to thank T. Y. Huang for his assistance
in SQUID magnetization measurements. The work at Binghamton was
partially supported by the SUNY-Bimghamton Research Foundation
(240-9522A).

\end{document}